# RF-induced evaporative cooling and BEC in a high magnetic field

P. BOUYER[1], V. BOYER[1], S.G. MURDOCH[1], G. DELANNOY[1],
Y. LE COQ[1], A. ASPECT[1] AND M. LÉCRIVAIN[2]

[1]Groupe d'Optique Atomique Laboratoire Charles Fabry de l'Institut d'Optique
UMRA 8501 du CNRS Orsay, France.
[2]L.E.Si.R. URA 1375 du CNRS - ENS Cachan, France

## 1. INTRODUCTION

Bose-Einstein condensates[1-3] are very promising for atom optics[4-6], where they are expected to play a role as important as lasers in photon optics, since they are coherent sources of atoms with a very large luminosity. In view of applications, it is crucial to develop apparatuses that produce BEC faster - the average production rate of a condensate is 0.01 Hz - and with more versatile designs, by reducing, for example, the power consumption of the electromagnets.

For this purpose, we have developed a magnetic trap for atoms based on an iron core electromagnet, in order to avoid the large currents, electric powers, and high pressure water cooling, required in schemes using simple coils. The latest developments allow us to achieve a very high confinement that will permit to achieve much higher production rates.

In this chapter, we will first present the design of the iron core electromagnet and how to solve the specific experimental problems raised by this technique. After presenting the experimental set-up, we will address the interruption of runaway evaporative cooling when the Zeeman effect is not negligible compared to the hyperfine structure. We will then present two ways to circumvent this problem: use of multiple RF frequencies and sympathetic cooling. Another method, hyperfine evaporation, was used in Ref. 2. In conclusion, we will present some applications of these high magnetic fields.

## 2.        IRON-CORE ELECTROMAGNET TRAP FOR ATOMS

Our iron-core electromagnet is shown Fig.2. It follows the scheme of Tollett *et al.*[7]. Instead of using permanent magnets, we use pure iron pole pieces excited by coils, which allows us to vary the trap configuration. The use of ferromagnetic materials was reported in Ref.8 .The role of the pole pieces is to guide the magnetic field created by the excitation coils far away from the center of the trap towards the tips of the poles. To understand this effect, let us consider the magnetic circuit represented Fig. 1.

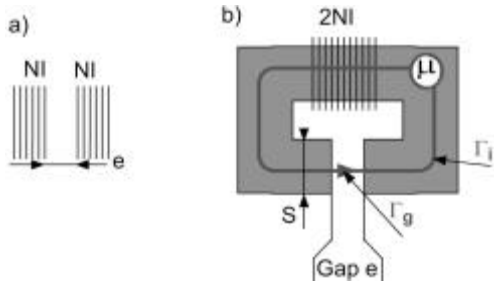

*Figure 1.* a) System of two anti-Helmholtz coils of excitation *NI* separated by *e*. b) Equivalent magnetic circuit. *l*: length of the ensemble poles + yoke. *e*: gap between the tips.

The two tips are separated by a gap *e* of a few centimeters. The ferromagnetic structure has a total length *l* and a section *S*. The whole structure is excited by a coil of *2N* loops driven by a current *I*, leading to an excitation *2NI*. From the Ampere's theorem[9], we can introduce the reluctance $R_{\mathrm{iron}}$:

$$R_{\mathrm{iron}} = \oint_{\Gamma_i} \frac{dl}{\boldsymbol{m}_r \boldsymbol{m}_0} \cong \frac{l}{\boldsymbol{m}_r \boldsymbol{m}_0} \tag{1}$$

inside the iron core and $R_{\mathrm{gap}}$:

$$R_{\mathrm{gap}} = \oint_{\Gamma_g} \frac{dl}{\boldsymbol{m}_0} \cong \frac{e}{\boldsymbol{m}_0} \tag{2}$$

in the gap between the tips. A simple relation between the excitation $2NI$ and the magnetic flux $BS$ can be written:

$$BS = \frac{2NI}{R_{\text{iron}} + R_{\text{gap}}}. \qquad (3)$$

In our case, the gap $e$ and the size of the ferromagnets $l$ are comparable. Since $\mathbf{m}$ is very important ($\mathbf{m} > 10^4$) for ferromagnetic materials, only the gap contribution is important[i]. A more complete calculation shows that the field created in the gap is similar to that created with two coils of excitation $NI$ placed close to the tips as represented[10] Fig. 1. Thus, guiding of the magnetic field created by arbitrary large coils far away from the rather small trapping volume is achieved. All this demonstration is only valid if a yoke links a north pole to a south pole. If not, no guiding occurs and the field in the gap is significantly reduced.

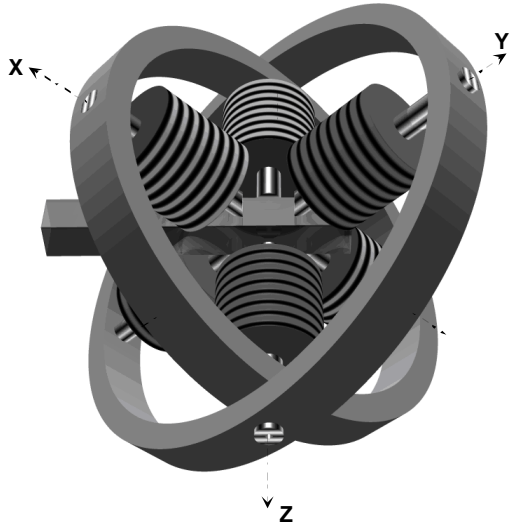

*Figure 2.* Diagram showing the position of the pole pieces of the electromagnet. The tip-to-tip spacing is 3 cm and the section of the cell is $1 \times 2$ cm$^2$.

We will focus now on our Ioffe-type trap for Rubidium 87 (Fig. 2), which consists of a superposition of a linear quadrupole field and dipole field. The linear quadrupole field of gradient $G$ is produced by two pairs of coils in an anti-Helmholtz configuration, along the $x$- and $y$-axis, and the dipole field of curvature $C$ is produced by two coils along the $z$-axis in a Helmholtz configuration[8]. The magnitude of the total magnetic field can be approximated by:

$$|\mathbf{B}| = B_0 + \left(\frac{G^2}{2B_0} - \frac{C}{2}\right)\left(x^2 + y^2\right) + Cz^2 \tag{4}$$

leading to an anisotropic axial harmonic potential for trapping states, in the linear Zeeman effect regime.

---

[i] The case of very small gap where $R_{\text{iron}} > R_{\text{gap}}$ was studied in Ref.[10]. In this case, the ferromagnetic materials amplify the magnetic field in the gap.

The use of ferromagnetic materials raises several problems:

**Geometry.** As mentioned previously, a ferromagnetic yoke has to link a north pole to a south pole. A bad coupling between two poles can result in reduced performances of the electromagnet. Our solution shown Fig. 2 optimizes the optical access around the vacuum cell while keeping the required coupling efficient.

In addition, the geometry of the magnetic field relies on the shape of the pole pieces rather than the geometry of the exciting coils. This results in the fact that the bias field $B_0$ cannot be easily decreased without also canceling the dipole curvature[ii] $C$. This implies a high bias field, for which the Zeeman shift is no longer linear in the magnetic field, due to a contamination between hyperfine levels. In fact, a more complicated design of the poles along the $z$-axis allows for canceling the bias field while keeping an important dipole curvature. This new design will be discussed in the last section of this chapter.

**Hysteresis.** Hysteresis prevents from returning to zero magnetic field after having switched ON and OFF the electromagnet. A remanent field of a few Gauss remains, as shown Fig. 3.

This remanent magnetization needs to be cancelled in order, for example, to release the atoms and make a velocity (temperature) time-of-flight (TOF) measurement. Extra coils around the magnetic poles (see Fig. 4), carrying a DC current will shift the hysteresis cycle so that it crosses zero again. The current is adjusted to provide the coercive excitation which cancel exactly the remanent magnetic field when the large coils are switched off. This compensation is valid as long as we remain on the same excitation cycle. This stability is achieved thanks to a computer control of the experiment.

---

[ii] In the systems using coils, an additional compensation coils is used the reduce the bias field. In our case, this additional external excitation would couple into the ferromagnetic structure, decreasing both the bias field and the curvature of the field.

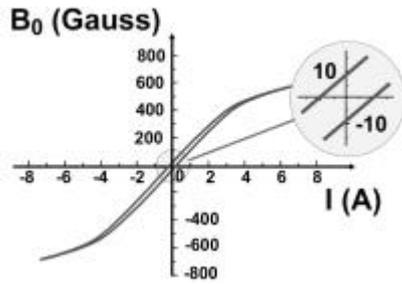

*Figure 3*. Hysteresis cycle of a ferromagnetic structure. The insert shows the typical remanent field value before compensation.

**Dynamic properties.** The use of big coils (lots of loops) results in a big inductance ($\cong$ 100 mH), leading to a switching time $\tau = L/R \cong$ 100 ms, too long to allow a good transfer of atoms into the magnetic trap. By assisting the switching with a capacitor, we are able to reduce $\tau$ to less than a millisecond.

Eddy currents are expected to seriously slow down the switching, and indeed a field decay constant of more than 10 ms was found in our first electromagnet[8]. The use of laminated material (stacked 100 **m**m/1 mm thick layers of ferromagnetic materials isolated by epoxy) solves this problem and allows to switch ON or OFF the field within 100 **m**s.

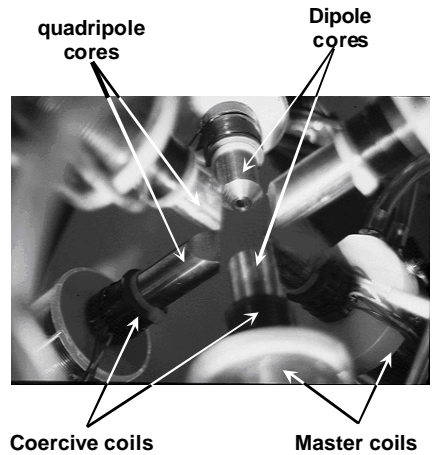

*Figure 4.* Photo of the poles, the excitation coils and the compensation coils.

## 3.        EXPERIMENTAL SETUP

The experimental setup is shown on Fig. 5. The electromagnet is placed around a glass cell of inner section of $1 \times 2$ cm$^2$, pumped with two ion pumps and a titanium sublimation pump. The background pressure is of the order of $10^{-11}$ mbar. The tip to tip spacing is 3 cm for the poles of the dipole, and 2 cm for the poles of both quadrupoles. The power consumption is 25 W per coil for a gradient of 900 Gauss/cm, and the maximum gradient at saturation is 1400 Gauss/cm.

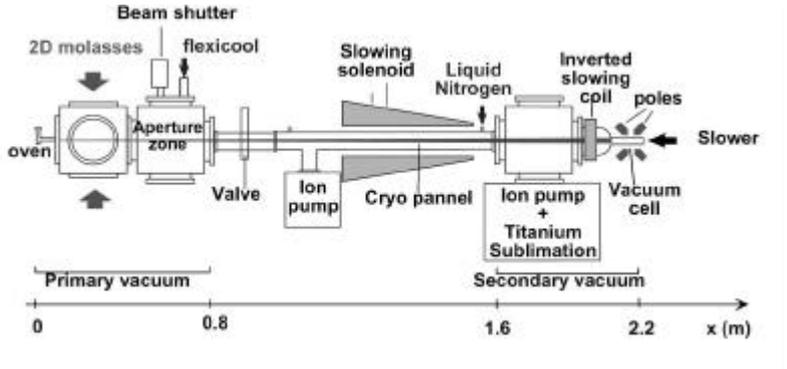

*Figure 5.* Experimental apparatus.

Our source of atoms is a Zeeman slowed atomic beam of $^{87}$Rb. The beam is collimated with a transverse molasses and is decelerated in a partially reversed solenoid. It allows us to load a MOT with $1.5 \times 10^9$ atoms in 5 s. In order to increase the density, we then switch to a forced dark MOT by suppressing the repumper in the center and adding a depumper tuned to the $F = 2 \rightarrow F' = 2$ transition[12,13]. We obtain $8 \times 10^8$ atoms at a density of $1.5 \times 10^{11}$ cm$^{-3}$. After additional molasses cooling, we optically pump the atoms into either the $F = 2$ or the $F = 1$ state. We then switch on the electromagnet in a configuration adapted to the phase space density of the atomic cloud. The bias field $B_0$ is fixed to $\sim 140$ Gauss for $F = 2$, or to $\sim 207$ Gauss for $F = 1$. The corresponding oscillation frequency is $\Omega/2\pi = 21$ Hz for $F = 2$ and $\Omega/2\pi = 18$ Hz for $F = 1$. We end up with $N = 4 \times 10^8$ trapped atoms at a temperature of 120 **n**K, with a peak density of $5 \times 10^{10}$ cm$^{-3}$. All this information is obtained by conventional absorption imaging on a CCD camera.

## 4.     INTERRUPTED EVAPORATIVE COOLING IN A HIGH MAGNETIC FIELD

To achieve Bose-Einstein Condensation (BEC), we use RF-induced evaporative cooling of the $^{87}$Rb atoms confined in the magnetic trap.

In the approximation of the magnetic moment of the atom adiabatically following the direction of the field during the atomic motion, the magnetic potential is a function of the modulus of the field and the projection $hm$ of the total angular momentum on the field axis. Depending on the sign of $m$, the Zeeman sublevel will be confined towards (trapping state) or expelled from (non-trapping state) a local minimum of the field modulus. RF-induced evaporative cooling consists of coupling the trapping state to a non-trapping state with a radio-frequency field (RF knife), in order to remove the most energetic atoms from the trap.

Efficient evaporative cooling[14-16] relies on fast thermal relaxation, and thus on the ability of increasing the collision rate by adiabatic compression of the atomic cloud. The most widely used mean to increase the curvature of the trapping potential is to partially cancel the bias field $B_0$ with two additional coils in Helmholtz configuration along the $z$-axis. As seen in Eq.(4), this increases the radial curvature without changing the axial curvature. Typical values of the compensated bias field in previous experiments are 1 to 10 Gauss. One can also radially compress the atomic cloud by increasing the gradient $G$ without modifying the bias field. This is the approach for our trap. However, the quadratic Zeeman effect is not negligible anymore.

Defining

$$\boldsymbol{e} = \frac{\boldsymbol{m}_B B_0}{2\hbar \boldsymbol{w}_{\mathrm{hf}}} = \frac{\boldsymbol{w}_{\mathrm{L}}}{\boldsymbol{w}_{\mathrm{hf}}} \tag{5}$$

as the ratio of the linear Zeeman effect $?_{\mathrm{L}}$ o the hyperfine splitting of the ground state $?_{\mathrm{hf}}$, we may write the Zeeman effect for $^{87}$Rb to the second order in e as

$$E_{F,m_F}(\boldsymbol{e}) - E_F = (-1)^F \left\{ \hbar \boldsymbol{w}_{\mathrm{hf}} m_F \boldsymbol{e} + \hbar \boldsymbol{w}_{\mathrm{hf}} \left(4 - m_F^2\right) \boldsymbol{e}^2 \right\} \tag{6}$$

with

$$E_F = \frac{\hbar \boldsymbol{w}_{\mathrm{hf}}}{2} \left(1 + (-1)^F\right) \tag{7}$$

the hyperfine structure. It gives a quadratic Zeeman effect of about 2 MHz at 100 Gauss (a typical bias field for our trap). An immediate consequence is that the ($F = 2$, $m_F = 0$) state is a trapping state. For a small magnetic field, as in most experiments, the second order in Eq.(6) is negligible. The RF coupling between adjacent Zeeman sublevels ( $|\Delta_m| = 1$) results in an adiabatic multiphoton transition to a non-trapping state, leading to efficient RF induced evaporative cooling. In the case of a high bias field, the RF couplings are not resonant at the same location because of the quadratic Zeeman effect. Depending on the hyperfine level, this effect leads to different scenarios[17], that we have experimentally identified[18] thanks to our magnetic trap allowing strong confinement with a high bias field.

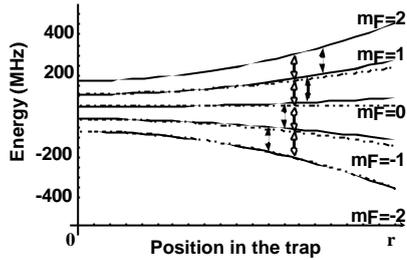

*Figure 6.* Forced evaporative cooling in the presence of the quadratic Zeeman shift. We have plotted the magnetic energy of the ($F = 2$, $m_F = 0$) sublevels as a function of a trap coordinate $r$ (the magnetic field $B(r) = B_0 + b \ r^2$). For a given RF frequency, the various transitions do not happen at the same position. We have indicated with black arrows the resonant transitions and with white arrows what would happen in the absence of quadratic corrections.

For atoms in the ($F = 2$, $m_F = 2$) state, forced evaporative cooling will be subject to unwanted effects as the RF knife gets close to the bottom of the potential well. Indeed, we can only cool down the sample to about 50 $\textbf{\textit{m}}$K until the atoms cannot be transferred to a non-trapping state and cooling stops. In addition, a careful analysis of the evaporation shows that it can only be optimized to rather poor efficiency. In order to give an insight of the efficiency of the forced evaporation in such a situation, let us consider an atom initially in the $m_F = 2$ trapping state, and following the path represented in Fig. 7 to connect to the $m_F = -1$ non-trapping state.

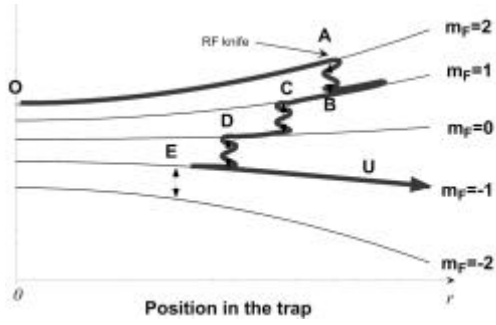

*Figure 7.* One short paths connecting the trapping state *(F = 2, $m_F$ = 2)* to the non-trapping state *(F = 2, $m_F$ = −1)*. The atom crosses 5 times the RF knife without making a transition at point **B** and **E**.

The atom, travelling from the center of the trap, reaches the RF knife at **A**, and makes a transition to the $m_F = +1$ state at **B** with a transition probability *P*. From there, it continues to move away from the center. When it comes back towards the center of the trap, the atom passing on **B** must not make a transition in order to reach the RF knife on **C**. The probability to reach **BC** from **OA** is *P*(1 - *P*). Assuming the same probability *P* for all the RF transitions, the probability that the atoms follows the path shown in Fig. 7 and leaves the trap on **EU** is $P^3(1 - P)^2$. There are 4 analogous paths involving 5 crossings of the RF knife. Consequently, neglecting interference effects, the total probability associated to these 4 short evaporation paths is $4P^3 (1 - P)^2$. This probability has a maximum value of about 10% for a transition probability *P* of $\sqrt{3}/5$, and is associated to a precise value of the atomic velocity. When considering all possible velocities, the probability of leaving the trap on $m_F = -1$ averages to less than 10%, much less than for the standard situation where the adiabatic passage has 100% efficiency for almost all velocities[17,19]. The experimental observation on Fig. 8 supports this simple analysis: when we increase the RF power, the efficiency of the evaporation reaches a maximum and then decrease[iii].

In addition, all paths longer than the 5 crossings path as in Fig. 7 contribute to build up a macroscopic population in the intermediate levels $m_F = 1$ and $m_F = 0$, as soon as evaporation starts. This results in the presence of the atoms intermediate sublevels during the evaporation and the observation of a heating of 5 **n**K/s of the $m_F = 2$ cloud, when we remove the RF knife at the end of the evaporation[19].

---

[iii] Of course, with sufficient RF power (P > 100 W), we would eventually reach a situation where all the various transitions merge, and a direct adiabatic transition to a non-trapping state with 100% efficiency would be obtained.

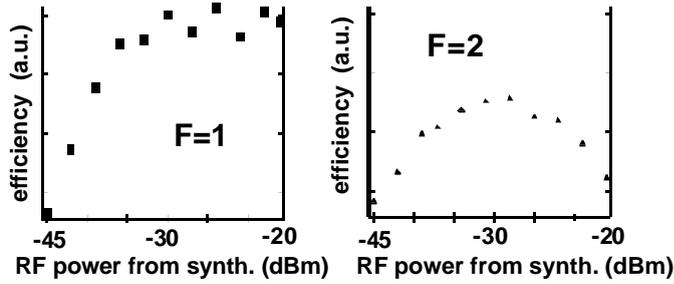

*Figure 8.* Efficiency of the evaporation for $F = 1$ (left) and $F = 2$ (right) as observed via the optical thickness of the atomic cloud. A clear optimum can be observed in the $F = 2$ case (the relative height between $F = 1$ and $F = 2$ is arbitrary).

Forced evaporative cooling for atoms trapped in ($F = 1$, $m_F = -1$) is not adversely affected by the quadratic Zeeman effect at a bias field $B_0$ of 207 Gauss, since the $m_F = 0$ state is non-trapping because of the sign of the quadratic term. The RF power has to be large enough to ensure an adiabatic transfer to $m_F = 0$ with an efficiency close to 1. For atoms in the F = 1 state, after adiabatic compression, the oscillation frequencies are $\Omega_\parallel/2\pi = 18$ Hz along the dipole axis, and $\Omega_\perp/2\pi = 55$ Hz along both quadrupole axis. We could successfully cool down the sample, and we obtained a condensate of a few $10^6$ atoms as shown Fig. 9.

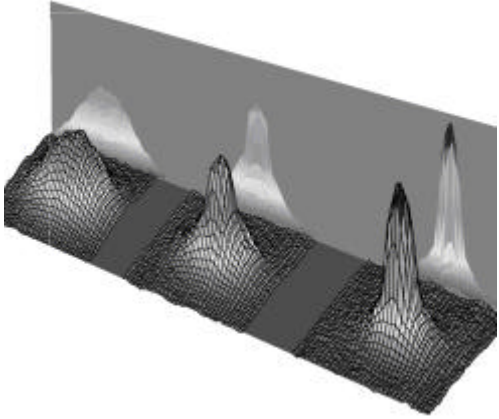

*Figure 9.* Bose Einstein condensation in $F = 1$, $m_F = -1$

# 5.    REACHING BEC IN F=2 IN HIGH MAGNETIC FIELD

Several strategies can circumvent the adverse consequences of the quadratic Zeeman effect, and achieve efficient forced evaporative cooling of $^{87}$Rb in $F = 2$.

## 5.1    Evaporation with 3 RF knives[22]

When evaporating the ($F = 2$, $m_F = 2$) state of $^{87}$Rb in a high bias field trap like ours, the RF couplings between the adjacent magnetic sublevels are not resonant at the location in the trap. Thus the transfer of atoms from trapping to non-trapping states is inefficient (or even non-existent). This problem can be overcome if we evaporate with three distinct RF frequencies chosen so that a direct transition to a non trapping state is always possible, as shown in Fig. 10.

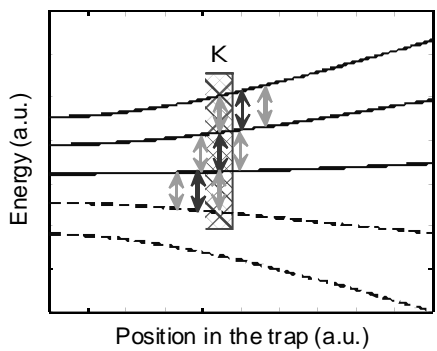

*Figure 10.* Implementation of 3 RF knives to evaporate in a high magnetic field. All possible transitions are represented. Frequency matching is only achieved at the RF knife *K*.

However the requirement of three independent frequency sources is not very practical[iv]. Rather a simpler solution involving the mixing of two frequencies will compensate the quadratic term of the Zeeman shift. As calculated on Eq.(6), the difference between successive RF transition is the same. Thus, compensation is achievable by mixing one independent frequency source (the carrier) with another RF frequency source of frequency $d\nu$ to obtain a pair of sidebands of fixed detuning and

---

[iv] Apart from the technological complexity, the mixing of 3 different frequencies can generate sidebands that will induce stray RF knives, reducing the evaporation efficiency.

approximately the same power as the carrier[v]. This detuning by is chosen as to align the three knives perfectly at the end of the evaporation ramp.

This approach will be limited to magnetic fields where the higher order Zeeman terms are not significant[vi]. Indeed, one has to compare the quadratic correction in Eq.(6) with the well-known Breit-Rabi formula:

$$E_{F,m_F}(B) - E_F = -m_F g_I m_n B + (-1)^F \frac{\hbar w_{hf}}{2} \left( \sqrt{1 + m_F x + x^2} - 1 \right) \quad (8)$$

with

$$x = \left( g_s m_B - g_I m_n \right) \frac{B}{\hbar w_{hf}} \cong 4e \qquad (9)$$

where $g_S$ and $g_1$ are the electronic and nuclear g-factors, and $m_n$ the nuclear magneton. The RF frequencies between the sublevels calculated from Eq.(8) are shown in table 1. We list only the transitions required to transfer the atoms to the first non-trapping state ($F = 2$, $m_F = -1$).

*Table 1.* Zeeman effect without approximations for different bias fields

| B | $|2,2\rangle \rightarrow |2,1\rangle$ | $|2,1\rangle \rightarrow |2,0\rangle$ | $|2,0\rangle \rightarrow |2,-1\rangle$ |
|---|---|---|---|
| 56 (Gauss) | 39.08-0.44 (MHz) | 39.08 (MHz) | 39.08+0.45 (MHz) |
| 111 | 77.12-1.66 | 77.12 | 77.12-1.78 |
| 207 | 142.18-5.43 | 142.18 | 142.18+6.12 |

From table 1 we can immediately see that for a bias field of 207 Gauss this approach will not work : it is impossible to choose a sideband detuning $dv$ for which either the *(F = 2, $m_F$ = 2) → (F = 2, $m_F$ = 1)* or the *(F = 2, $m_F$ = 0) → (F = 2, $m_F$ = −1)* transition will not be detuned from resonance by at least 500 kHz. This is much larger than the available RF power broadening estimated to be of about 10 kHz. Indeed, experimentally when evaporating *(F = 2, $m_F$ = 2)* with the 3 RF knifes in this bias field we are unable to cool the atoms below 15 $m$K. However this is an order of magnitude lower than the lowest temperature we can obtain when evaporating *(F = 2, $m_F$ = 2)* with only one RF knife (100 $m$K). In a bias field of 111 Gauss the situation is already better with a optimum detuning of the two sidebands from their respective resonance of 50 kHz. Here we can cool the ($F = 2$, $m_F = 2$) cloud down to 500 nK, and obtain a phase space density of 0.1. We believe that with just a little more RF power or better initial conditions for the evaporation the condensation of ($F = 2$, $m_F = 2$) should be possible with this technique for this bias field. When we again lower the bias field by a factor of two to 56 Gauss the effect of the nonlinear terms of the Zeeman shift higher than the quadratic correction becomes negligible compared to the RF power broadening. Here we were able to cool atoms below 100 nK and could attain BEC in ($F = 2$, $m_F = 2$) as desired. It should be noted that the effect of the quadratic correction to the Zeeman shift is significant here, since with one RF knife we are unable to cool the cloud below 10 *n*K. Figure 11 shows a graph of the measured number of atoms in a condensate of ($F = 2$, $m_F = 2$) as a function of the sideband detuning $\boldsymbol{d}v$. The optimal detuning of the RF sidebands from the central carrier is measured to be 0.45 MHz in good agreement with the prediction of table 1. The width of the curve in Fig. 11 is in good agreement with the estimated Rabi frequency and with the residuals calculated with Eq.(8). From this, we can conclude that the average Rabi frequency of our RF knives is indeed of the order of 10 kHz.

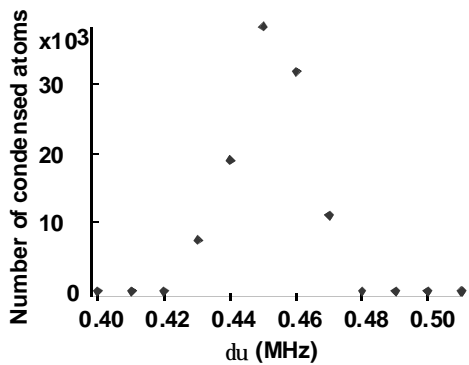

*Figure 11* Bose Einstein condensation with 3 RF knives : number of atoms in the condensate versus $\boldsymbol{d}v$.

## 5.2    Sympathetic Cooling

Another possible path to condensation in the ($F = 2$, $m_F = 2$) state of $^{87}$Rb in a high magnetic field is to use sympathetic cooling[20]. In sympathetic cooling, one evaporatively cools one species of atom, a second species being cooled simply by thermal contact with the former. In our case, we evaporate ($F = 1$, $m_F = -1$) which we know may be efficiently cooled by the standard method, even in a high magnetic field, and use them to cool atoms in ($F = 2$, $m_F = 2$). This cooling method is nearly lossless for the ($F = 2$, $m_F = 2$) atoms as they are evaporated in a potential twice as strong as the ($F = 1$, $m_F = -1$) atoms.

The efficiency of sympathetic cooling can be estimated with a simple model assuming that the two species are always at thermal equilibrium. The total energy of the system can be written

$$E = 3N_1 k_B T + 3N_2 k_B T \; .$$
(10)

The energy taken away by $dN_1$ atoms evaporated at height $\pmb{h}$[14-16] is

$$dE = 3dN_1 (\pmb{h} - 1) k_B T$$
(11)

and the energy of the atoms remaining trapped after evaporation of these $dN_1$ atoms is

$$E - dE = 3(N_1 - dN_1) k_B (T - dT) + 3N_2 k_B (T - dT)$$
(12)

since the number $N_2$ of atoms in ($F = 2$, $m_F = 2$) is nearly constant during evaporation. If we assume that evaporation is performed at fixed height $\pmb{h}$, one can simply integrate Eq.(12) by replacing $E$ and $dE$ by their expression in Eqs. (10) and (11). This results in the equation

$$\frac{T^f}{T^i} = \left( \frac{N_1^f + N_2}{N_1^i + N_2} \right)^{\frac{(\pmb{h}-2)}{3}}$$
(13)

relating the ratio between initial temperature $T^i$ and final temperature $T^f$ with the loss of atoms in ($F = 1$, $m_F = -1$).

For example, if we choose $\pmb{h}$ to be 5 (a typical value for experiments) and if we suppose that $N_2 \ll N_1^i$ one can immediately see that the minimum achievable temperature $T_{min}$ scales as the initial ratio $N_2 / N_1^i$. We can now estimate if the initial conditions are sufficient to achieve BEC. For that, we need to compare $T_{min}$ to the critical temperature

$$T_c(F = 1) = \sqrt[3]{\frac{N_1^f}{1.202}} \left( \hbar \sqrt{\frac{\pmb{m}_B G^2}{2 B_0 M}} \right) = K \sqrt[3]{N_1^f}$$
(14a)

$$T_c(F = 2) = \sqrt{2} \; K \sqrt[3]{N_2}$$
(14b)

for each of the 2 species. We can easily see that the initial ratio $N_2 / N_1^i$ can be chosen to either condense ($F = 1$, $m_F = -1$) before ($F = 2$, $m_F = 2$), or ($F = 2$, $m_F = 2$) before ($F = 1$, $m_F = -1$). If $N_2$ is too large, no condensation is possible and if $N_2$ is too small, only the ($F = 1$, $m_F = -1$) atoms can be condensed. This happens for a critical number of atoms $N_2^c$

$$N_2^c \cong \left( \frac{2\sqrt{2} K^3 \left(N_1^i\right)^{h-2}}{T_i^3} \right)^{1/(h-3)}.$$ (15)

In order to keep evaporative cooling efficient all the way towards BEC, one has to insure that the atoms remain in good thermal contact. Because of gravity the ($F = 1$, $m_F = -1$) cloud is centered below the ($F = 2$, $m_F = 2$) cloud as it is more weakly trapped. This displacement between the two clouds is given by

$$D = \frac{Mg}{m_B} \frac{B_0}{G^2}.$$ (16)

For a fixed gradient G = 900 Gauss/cm this gives a variation with $B_0$, D = 0.16 *m*m/Gauss. The RMS width of a thermal cloud decreases with the square root of the cloud temperature during the cooling, so assuming the two clouds must be within one RMS width of each other for good thermal contact we can obtain an estimate for the minimum temperature to which ($F = 2$, $m_F = 2$) can be sympathetically cooled by atoms in ($F = 1$, $m_F = -1$), namely,

$$T_{\text{limit}} = \frac{M^2 g^2}{24 k_B m_B G^2} B_0$$ (17)

proportional to the bias field $B_0$. For a gradient G = 900 Gauss/cm (Eq. (17)) may be evaluated to give a variation with $B_0$ of $T_{\text{limit}} \cong 1.4$ nK/Gauss. The physical interpretation of this is simple; for our trap, the higher the value of the bias field, the weaker the confinement of the quadrupole, the larger the displacement between the two species and hence the higher the minimum possible temperature. Experimentally this simple theory was in good agreement with our observations. By a careful adjustment in the optical pumping cycle during the transfer to the magnetic trap we could start the evaporation with a small but controllable fraction of the atoms in the ($F = 2$, $m_F = 2$) state and the rest in the ($F = 1$, $m_F = -1$) state. For a bias field of 207

Gauss we found we were able to cool the atoms in ($F = 2$, $m_F = 2$) down to a temperature of 400 nK, in rough agreement with the simple estimate of Eq. (17) of 290 nK. The phase space density for the ($F = 2$, $m_F = 2$) cloud at this point was 0.05. Further cooling the ($F = 1$, $m_F = -1$) cloud did not reduce the temperature of the atoms in ($F = 2$, $m_F = 2$).

When we repeated this experiment for a bias field of 56 Gauss, we were able to condense ($F = 2$, $m_F = 2$) sympathetically in the presence of ($F = 2$, $m_F = -1$) for a sufficiently small initial number of atoms in ($F = 2$, $m_F = 2$). When the proportion of atoms in the ($F = 2$, $m_F = 2$) state is too large their rethermalization heats the cooling atoms in ($F = 1$, $m_F = -1$) too much for an efficient evaporation. Fig. 12 shows phase space density in each state as a function of the final frequency of the evaporation ramp, for three different initial numbers of atoms in the ($F = 2$, $m_F = 2$) state.

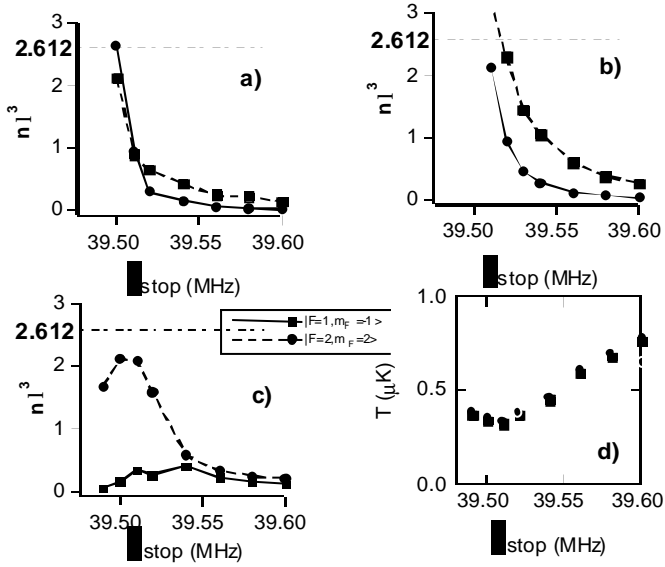

*Figure 12.* Sympathetic cooling representation of the phase space density of both species as a function of the RF frequency. **a)** The number of atoms in ($F = 2$, $m_F = 2$) is small enough so as not to significantly affect the evaporation of the atoms in ($F = 1$, $m_F = -1$): in the total absence of atoms in ($F = 2$, $m_F = 2$) we could produce condensate in ($F = 1$, $m_F = -1$) with a similar transition temperature and number of atoms. In this case the cloud of atoms in ($F = 1$, $m_F = -1$) attains sufficient phase space density $n\lambda^3$ to condense before the atoms in ($F = 2$, $m_F = 2$). **b)** The number of atoms in ($F = 2$, $m_F = 2$) is two times higher and the evaporation of the ($F = 1$, $m_F = -1$) cloud is significantly hampered by the sympathetic cooling of ($F = 2$, $m_F = 2$) atoms. Here the atoms in ($F = 2$, $m_F = 2$) attain sufficient phase space density to condense before the ($F = 1$, $m_F = -1$), as even though there are less of them they are more tightly confined in the magnetic trap. **c)** The number of atoms in ($F = 2$, $m_F = 2$) is too large too be cooled sympathetically to the temperature required for condensation **d)** Diagram showing the 2 clouds remain in good thermal equilibrium during all the evaporation.

## 6. AN APPLICATION OF HIGH BIAS FIELD: COUPLING BETWEEN 2 POTENTIAL WELLS

The quadratic Zeeman effect can be an asset rather than a nuisance once condensation is reached. For instance, one can make a selective transfer of part of the condensate from the ($F = 1$, $m_F = -1$) state to the ($F = 2$, $m_F = 0$) state by using a 6.8 GHz pulse. Thanks to the quadratic Zeeman effect, the ($F = 2$, $m_F = 0$) state is a very shallow trapping state (for a bias field of 56 Gauss, the oscillation frequencies are $\Omega_\perp^0 \leq 40\,\mathrm{Hz}$ along the quadrupole and $\Omega_z^0 \leq 10\,\mathrm{Hz}$ along the dipole), some features of a trapped Bose gas can eventually be observed more easily. We studied the weak coupling between those two states by turning on a weak 6.8 GHz RF knife.

The two-coupled potential wells are represented on Fig. 13. Because of gravity, the centers of these two harmonic traps are displaced by typically of 300 $\mu$m.

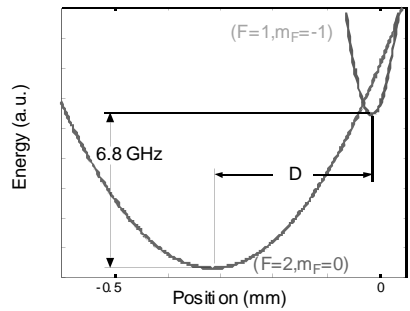

*Figure 13.* Energy diagram of the ($F = 1$, $m_F = -1$) and ($F = 2$, $m_F = 0$) states

We start with a condensate in the ($F = 1$, $m_F = -1$) state. We will restrict ourselves to the vertical dimension, where the two traps are offset. In the Thomas-Fermi approximation, this condensate is described by the wave function

$$\left| \psi(x) \right|^2 = \frac{\mu - U_1(x)}{\tilde{U}} \tag{18}$$

where $\mu$ is the chemical potential, $a$ the scattering length and $U_1(x)$ the trapping potential for the atoms in $F = 1$. We define

$$\tilde{U} = \frac{4\pi \hbar^2 a}{M}. \tag{19}$$

This state has a size of typically $s_x \cong 10\mu m$. The origin of coordinates is taken at the center of this trap. We will neglect the interactions in the ($F = 2$, $m_F = 0$) potential well. Consequently, the weak RF knives will couple the wave function $j\,(x)$ to the eigenstates $\Psi_n(x - D)$ of the $F = 2$ trapping potential. In good approximation, this potential is that of a harmonic oscillator of oscillation frequency $\Omega_\perp^0$ offset down by $D$ from the Bose-Einstein condensate. Thus, we can write

$$\Psi_n(x - D) = \frac{1}{2^n n!}\left(\frac{b^2}{p}\right)^{1/4} e^{-\frac{b^2 x^2}{2}} H^n(bx) \qquad (20)$$

where $H^n(x)$ is a Hermite Polynomial of order $n$ and

$$b = \sqrt{\frac{m\Omega_\perp^0}{\hbar}} \qquad (21)$$

a scaling parameter. The size of these eigenstates is given by

$$\Delta x_n = \frac{\sqrt{(n+1/2)}}{b}. \qquad (22)$$

The coupling efficiency is proportional to the overlap integral between $j\,(x)$ and $\Psi_n\,(x)$. Roughly, only the eigenstates $n$ such as $D\text{-}s_x \leq \Delta x_n/2 \leq D+s_x$ will be efficiently coupled. Since the condensate is highly coherent, the resulting wavefunction will be the coherent sum of those coupled eigenstates[vii]. This allows us to evaluate the atomic density created in the ($F = 2$, $m_F = 0$) state

$$\Pi(x) \cong \left|\sum_{n\backslash\Delta x_n \cong D \pm s_x} \Psi_n(x)\right|^2 \qquad (23)$$

as shown in Fig. 14. One clearly sees beatnotes between the different atomic modes. On the contrary, in the case of a thermal cloud of $F = 1$ atoms with approximately the same size, the resulting density distribution in the $F = 2$ trap will be the sum of the single eigenstates density profiles, since the coupled eigenstates will incoherently add-up.

---

[vii] Experimental studies[21] showed that the trapped condensate is coherent over its full length.

$$\Pi(x) \cong \sum_{n \, \setminus \Delta x_n \cong D \pm s_s} \left| \Psi_n(x) \right|^2 \tag{24}$$

this results in the disappearance of the periodic structure.

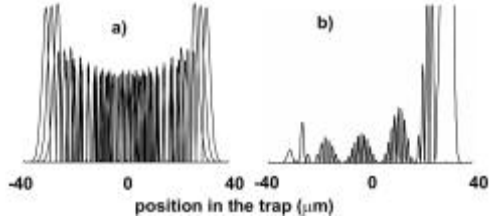

*Figure 14.* Simple picture of the effect of a weak coupling between the ($F = 1$, $m_F = -1$) state and the ($F = 2$, $m_F = 0$) state. Left, incoherent superposition of different eigenstates of the ($F = 2$, $m_F = 0$) trapping potential. Right, Atomic density profile $\Pi(x)$ of the coherent sum of the coupled eigenstates.

A more complete analysis can be done by computing a numerical solution of the coupled Gross-Pitaevskii equations.

$$i\hbar \frac{d}{dt}\boldsymbol{j}_1 = -\frac{\hbar^2}{2M}\nabla^2 \boldsymbol{j}_1 + U\boldsymbol{j}_1 + \tilde{U}\left(\left|\boldsymbol{j}_1\right|^2 + \left|\boldsymbol{j}_2\right|^2\right)\boldsymbol{j}_1 + \frac{\hbar\Omega_{RF}}{2}\boldsymbol{j}_2 \tag{25a}$$

$$i\hbar \frac{d}{dt}\boldsymbol{j}_2 = -\frac{\hbar^2}{2M}\nabla^2 \boldsymbol{j}_2 + U\boldsymbol{j}_2 + \tilde{U}\left(\left|\boldsymbol{j}_1\right|^2 + \left|\boldsymbol{j}_2\right|^2\right)\boldsymbol{j}_2 + \frac{\hbar\Omega_{RF}}{2}\boldsymbol{j}_1 . \tag{25b}$$

For simplicity, we supposed that the scattering length is the same for any binary elastic collision. A comparison of the numerical calculation and of preliminary experimental results is shown Fig. 15.

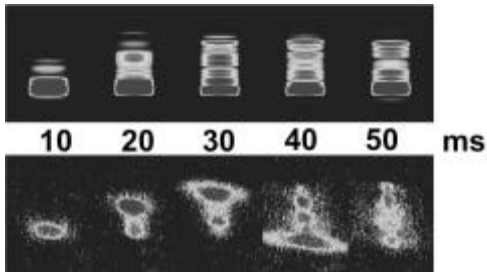

*Figure 15.* Effect of a weak coupling between the ($F = 1$, $m_F = -1$) state and the ($F = 2$, $m_F = 0$) state. Top, result of a numerical integration of the coupled Gross-Pitaevskii equation. The image takes into account corrections from the resolution of the imaging system and the camera. Bottom, experimental result of the weak coupling versus the time after the coupling was switched ON. One can clearly see the apparition of beatnotes between the atomic mode. With a cold but thermal cloud, no similar feature could be observed.

# 7.     IMPROVED IRON CORE ELECTROMAGNET TRAPS

A new design of the pole pieces allow for a compensated bias field $B_0$ - on the order of 1 Gauss - while keeping a significant value for the dipole curvature $C$ - on the order of 100 Gauss/cm$^2$. This, combined with an improved quadrupole gradient to 2400 Gauss/cm allows for a very high compression ratio. Depending on the initial number of atoms, this would allow to reach BEC in a few seconds.

The parameters of this new trap will also allow for studying new properties of BEC. Given a bias field of 80mG, this trap has a transverse field curvature of $7 \times 10^7$ G/cm$^2$, such that the ratio of the transverse to longitudinal field curvatures is $10^6 : 1$. This large asymmetry in the trapping potential will allow to form a 1D system. When the temperature of the system is low enough, particles are frozen into the quantum mechanical ground state of the transverse dimensions. However, since the ground state energy in the longitudinal direction is roughly $10^3$ times smaller than that of the transverse direction (since ground state energy scales as the square root of the field curvature), excited longitudinal states can still be occupied. In this one-dimensional regime, the physics of collisions, thermalization, and quantum degeneracy follow laws which are qualitatively different from those of the typical three-dimensional system.

## ACKNOWLEDGMENTS

This work is supported by CNRS, MENRT, Région Ile de France and the European Community. SM acknowledges support from Ministère des Affaires Étrangères.